\documentstyle[prb,aps,twocolumn]{revtex}
\begin{document}

\title{Large Scale Electronic Structure Calculations\\
with Multigrid Acceleration}

\author{E.\ L.\ Briggs, D.\ J.\ Sullivan, and J.\ Bernholc}
\address{Department of Physics,
North Carolina State University,
Raleigh, North Carolina 27695-8202}

\maketitle

\begin{abstract}
We have developed a set of techniques for performing large scale {\em
ab initio\/} calculations using multigrid accelerations and a
real-space grid as a basis.  The multigrid methods permit efficient
calculations on ill-conditioned systems with long length scales or
high energy cutoffs.  The technique has been applied to systems
containing up to 100 atoms, including a highly elongated diamond cell,
an isolated C$_{60}$ molecule, and a 32-atom cell of GaN with the Ga
d-states in valence.  The method is well suited for implementation on
both vector and massively parallel architectures.
\end{abstract}

\narrowtext

Over the course of the last several decades, plane-wave-based methods
have been used to perform electronic structure calculations on a wide
range of physical systems.  The Car-Parrinello (CP) and other
iterative methods
\cite{CarParrinello,PayneEtAl} have made such calculations possible for
systems containing several hundred atoms. \cite{BigCalculations} While
these methods have been very successful, several difficulties arise
when they are extended to physical systems with large length scales or
containing first-row or transition-metal atoms. Special techniques
have been developed to handle some of these problems.  Optimized
pseudopotentials,
\cite{Vanderbilt1990,RappeEtAl,LinEtAl,LiRabii} the augmented-wave method,
\cite{Bloechl} plane waves in adaptive-coordinates, \cite{Gygi1992} and
preconditioning combined with conjugate-gradient techniques
\cite{AriasEtAl,TeterPayneAllan} have had considerable success.  However,
these techniques are still constrained by the plane-wave basis set, which
requires periodic boundary conditions for every system and fast Fourier
transforms (FFTs) to efficiently transform between real and reciprocal
space. FFTs involve non-local operations that impose constraints on the
adaptability of these algorithms to massively parallel computer
architectures, because they perform best on problems that can be divided
into localized domains.

It has been appreciated for some time that there are potential
advantages to performing electronic-structure calculations entirely in
real space.  Boundary conditions are not constrained to be periodic,
which permits the use of non-periodic boundary conditions for clusters
and a combination of periodic and non-periodic boundary conditions for
surfaces.  By employing nonuniform real-space grids, it is possible to
add resolution locally; {\em e.g.}, for a surface or cluster
calculation, a basis that uses a high density of grid points in the
regions where the ions are located and a lower density of points in
the vacuum regions can lead to order of magnitude savings in the basis
size and total computational effort. \cite{BernholcYiSullivan} More
importantly, the use of a real space basis opens up the possibility of
using multigrid iterative techniques to obtain solutions of the
Kohn-Sham equations.  Multigrid methods \cite{Brandt} provide
automatic preconditioning on all length scales, which can greatly
reduce the number of iterations needed to converge the electronic
wavefunctions.  Furthermore, the real-space multigrid formulation does
not involve long-range operations and is particularly suitable for
parallelization and wavefunction-based $O(N)$ algorithms,
\cite{OrderN} because every operation can be partitioned into
hierarchical real-space domains.  In this Communication we describe
the essential elements of the uniform-grid, real-space multigrid
implementation as well as tests on a variety of systems, including a
vacancy in diamond, a strongly elongated 96-atom diamond supercell,
and a 32-atom supercell of GaN with 3d electrons in valence.

There have been a number of previous real-space grid-based
electronic-structure calculations.  The finite-element method was
applied by White {\em et al.\/} \cite{WhiteTeterWilkins} to
one-electron systems.  They used both conjugate-gradient and multigrid
acceleration to find the ground-state wavefunction.  Two of the
present authors \cite{BernholcYiSullivan} used nonuniform grids with
locally enhanced regions in conjunction with multigrid acceleration to
calculate the electronic properties of atomic and diatomic systems
with nearly singular pseudopotentials.  They verified that the
preconditioning afforded by multigrid was effective in
multi-length-scale systems.  Recently, Chelikowsky {\em et al.\/}
\cite{ChelikowskyEtAl} have used high-order finite-difference methods
and soft non-local pseudopotentials on uniform grids to calculate the
electronic structure, geometry, and short-time dynamics of small Si
clusters.

Several issues that are absent from plane-wave or orbital-based
methods arise when using a real-space grid approach.  In the former
case the wavefunctions, potentials, and the electronic density are
representable in explicit basis functions, and thus are known
everywhere.  Errors in the representation of these quantities are
mainly due to the truncation of the basis.  In a real-space grid
implementation, these quantities are known only at a discrete set of
grid points, which can introduce a spurious dependence of the
Kohn-Sham eigenvalues, the total energy, and the ionic forces on the
positions of the ions with respect to the real-space grid.  We have
developed a set of techniques that can overcome these difficulties and
can be used to compute accurate static and dynamical properties of
large physical systems, while taking advantage of the rapid
convergence rates that are made possible by multigrid methods.  In our
formalism the wavefunctions, density, and potentials are directly
represented on a uniform three-dimensional real-space grid with linear
spacing $h_{grid}$ and number of mesh points $N_{grid}$. The ions are
represented by soft-core norm-conserving
pseudopotentials. \cite{Hamann} Exchange and correlation effects are
treated using the local density approximation (LDA) of density
functional theory.

We start with the eigenvalue expression for the total electronic
energy of a system of electrons and ions in the local density
approximation (LDA):
\begin{eqnarray}
E_{LDA} =
&&
   \sum_n \epsilon_n + \int \, d\vec r \, \rho(\vec r) \,
   ( \epsilon_{XC}(\rho(\vec r)) - \mu_{XC}(\rho(\vec r)) ) \nonumber\\
&&
   -{1 \over 2} \int \, d\vec r \, \rho(\vec r) \, v_{Hartree}(\vec r)
   + E_{ion-ion}.
\label{LDA-Energy}
\end{eqnarray}
Since the kinetic energy operator is not explicitly represented in
Eq.\ (\ref{LDA-Energy}), we can discretize the Kohn-Sham equations in
a generalized eigenvalue formulation:
\begin{equation}
A[\psi_n] + B[V_{eff} \psi_n] = \epsilon_n B[\psi_n].
\label{compact-implicit}
\end{equation}
$A$ and $B$ are the components of the {\em Mehrstellen\/} discretization,
\cite{Collatz} which is based on Hermite's generalization of Taylor's
theorem.  It uses a weighted sum of the wavefunction and potential
values to improve accuracy of the discretization of the {\em entire\/}
differential equation, not just the kinetic energy operator.  The
weights are listed in Table \ref{Mehrstellen}.  The Kohn-Sham
hamiltonian, wavefunctions, and eigenvalues are represented to
$O(h_{grid}^4)$, but the prefactor for the $O(h_{grid}^4)$ term is
smaller than 0.01.  Alternative grid-based discretizations of the
Kohn-Sham equations do exist, {\em e.g.}, high-order
central-finite-difference methods were used by Chelikowsky {\em et
al.}\cite{ChelikowskyEtAl} They discretize the kinetic energy operator
only, and typically use an $O(h_{grid}^{12})$ discretization.  {\em
Mehrstellen} discretization differs from central-finite-difference
discretization in that it achieves higher accuracy by using more {\em
local} data.  For example, the fourth-order $A$ operator extends to
second nearest neighbors; in contrast, the twelfth-order
central-finite-difference operator extends linearly out 6 mesh points
in each direction.  We have found in our test cases that the
fourth-order {\em Mehrstellen\/} discretization produces equivalent or
better accuracy than the standard sixth-order finite-difference
discretization.

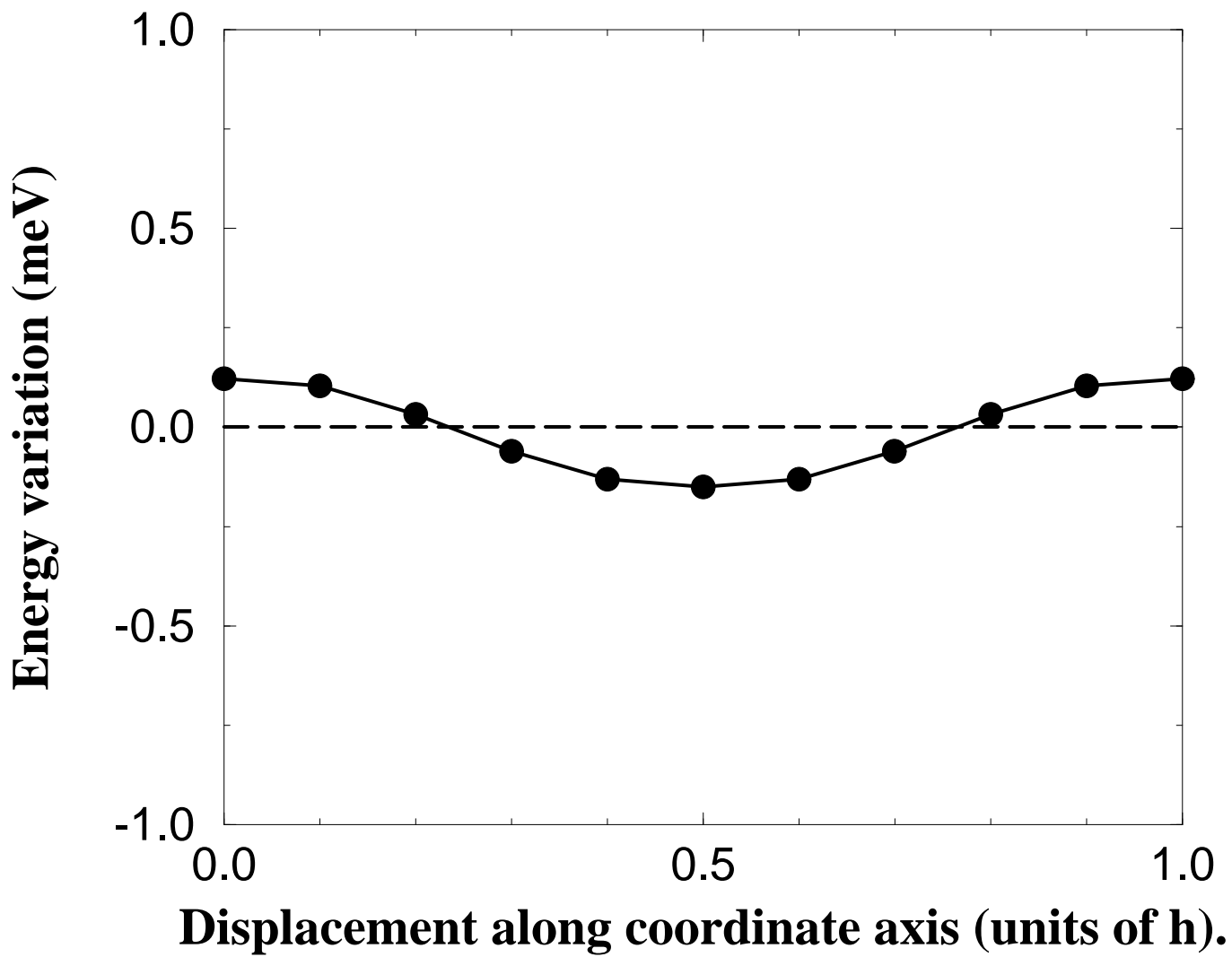
\begin{figure}
\caption{Variation of the total energy as a carbon atom is displaced
relative to the grid point along a coordinate axis.  The grid spacing
$h$ is 0.336 bohr.}
\label{CVariation}
\end{figure}

When using plane-wave basis sets, the accuracy is usually determined
by the convergence of the total energies and eigenvalues as the energy
cutoff used is increased.  In calculations employing a real-space
grid, it is also necessary to consider the dependence of the energies
and the ionic forces on the positions of the ions relative to the
real-space grid points.  This non-physical dependence is not present
in plane-wave methods.  Our selection of the {\em Mehrstellen\/}
operator and discretization techniques is designed to reduce this grid
dependence as much as possible, because the determination of accurate
trajectories in molecular dynamics simulations and the calculation of
dynamical quantities is very sensitive to this dependence. Our
approach is to perform calculations on isolated atoms and to monitor
the variations as the ion is moved relative to the grid points. To
reduce finite-size effects, periodic boundary conditions and a large
supercell are used in these tests rather than cluster boundary
conditions.  In Fig.\ \ref{CVariation} we show the variation in the
total energy of a carbon atom.  For bulk calculations, we also test
the grid resolution by monitoring the total energy as the atomic
lattice is rigidly translated with respect to the grid.

The solution of Eq.\ (\ref{compact-implicit}) by direct matrix methods
or by standard iterative techniques is prohibitively expensive for
very large systems because they require $O(N_{grid}^{5/3})$ operations
per wavefunction on an uniform mesh of $N_{grid}$ points. \cite{MGTut}
As the grid resolution is increased, {\em e.g.}, to treat systems
containing first row or transition metal atoms, the rate of
convergence becomes unacceptable.  A similar slow-down is present in
plane-wave methods as the energy resolution increases.

To efficiently solve Eq.\ (\ref{compact-implicit}), we have used
multigrid iteration techniques that accelerate convergence by
employing a sequence of grids of varying resolutions.  The final
solution is obtained on a grid fine enough to accurately represent the
pseudopotentials and the electronic wavefunctions.  If the solution
error is expanded in a Fourier series, it may be shown that iterations
on any given grid level will quickly reduce the components of the
error with wavelengths comparable to the grid spacing but are
ineffective in reducing the components with wavelengths large relative
to the grid spacing. \cite{Brandt,MGTut} The solution is to treat the
lower frequency components on a sequence of auxiliary grids with
progressively larger grid spacings, where errors appear as high
frequency components.  This procedure provides excellent
preconditioning for all length scales present in a system and leads to
very rapid convergence rates.  The operation count to converge one
wavefunction with a fixed potential is $O(N_{grid})$, compared to
$O(N_{grid}\,logN_{grid})$ for FFT-based approaches.
\cite{PayneEtAl}  In addition, all operations except orthogonalization are
short ranged, which allows for easy parallelization and a natural
implementation of $O(N)$ algorithms. \cite{OrderN}

One iteration towards the solution of Eq.\ (\ref{LDA-Energy}) consists
of a multigrid step to solve Eq.\ (\ref{compact-implicit}), followed
by orthogonalization of the orbitals, and an update of the electronic
density.  The components of the effective potential that depend on the
density (Hartree and exchange-correlation) are then recomputed for the
new density and the process is repeated until self-consistency is
reached.  We have found that it is often necessary to implement a
mixing procedure when updating the electronic density to minimize
charge sloshing. \cite{Pickett} The Hartree potential is computed by
solving Poisson's equation using multigrid iterations on the
corresponding {\em Mehrstellen\/} discretization.  The non-local ionic
pseudopotential is evaluated in real space using the Kleinman-Bylander
\cite{KleinmanBylander} separable form.  To facilitate comparisons
with plane-wave calculations, we define an equivalent energy cutoff
for the multigrid calculation, $\pi^2/2h^2$ [Ry], to be equal to that
of a plane-wave calculation that uses a FFT grid with the same spacing
as the multigrid calculation. \cite{Cutoff} Using this convention, the
computational time required to perform one step of the above procedure
is approximately the same as for a single step in the CP method when
the cutoffs are equal.

The multigrid method outlined above has been tested on several large
systems.  A 64-atom diamond supercell was chosen as a representative
periodic system for which both multigrid and Car-Parrinello
calculations are feasible.  The calculations were performed on the
perfect crystal and the relaxed vacancy using the same
pseudopotential. \cite{Hamann} The CP cutoff was 35 Ry, while the
multigrid code used a grid spacing of 0.336 bohr, which results in a
grid with the same total number of points as the FFT grid needed in
the CP code.  The k-space sampling was restricted to the $\Gamma$
point.  The dynamical relaxation method \cite{ZhangYiBernholc} was
used to quickly relax the vacancy in both calculations.

The Car-Parrinello and real-space calculations are in excellent
agreement with each other.  The relaxed ionic-positions agree to
within 0.009 bohr for all ions in the supercell, and the largest
difference in Kohn-Sham eigenvalue is 0.06 eV.  The diamond vacancy
introduces an initially triply degenerate level in the gap, which
splits into a singlet and a doublet due to the Jahn-Teller effect.
The magnitude of this splitting is 0.32 eV in both the real-space and
Car-Parrinello results.  The main results are summarized in Table
\ref{Diamond}.  The discrepancies between the LDA calculations and
experimental values for the bandgap and the cohesive energy are
well-known deficiencies of density functional theory.

An isolated C$_{60}$ molecule was selected as an example of a
non-periodic system.  The simulation cell was a cube of length 23 bohr
and the grid spacing was 0.360 bohr. The initial ionic coordinates
were generated using the classical Tersoff-Brenner potential,
\cite{Tersoff} and the electronic wavefunctions were set to zero on
the boundaries of the cell.  After the convergence of the electronic
system, the ions were relaxed using the same relaxation scheme as
before.  Two distinct bond lengths were found in the final structure,
corresponding to the carbon-carbon single and double bonds.  There
were twice as many single bonds as double bonds, and the average
double and single bond lengths were 1.39 and 1.44 \AA, compared to
1.41 and 1.45 \AA\ obtained in a previous CP calculation
\cite{ZhangYiBernholc} for the C$_{60}$ solid.  The standard deviations of the
bond-length distributions were on the order of $10^{-3}$ \AA\ in both
calculations. The experimental values for the solid are 1.40 and 1.45 \AA,
respectively.

A significant advantage of the multigrid method is the speed of
convergence to the electronic ground state for a given initial ionic
configuration.  For systems requiring a small energy cutoff or of
small size, the speed advantage with respect to CP-based methods is
not substantial. However, for systems requiring a large energy cutoff,
or of large dimensions, this advantage---as measured in actual
computational time---is typically an order of magnitude.  This is
because the maximum stable timestep in the Car-Parrinello method must
be much smaller than in the multigrid approach.

To illustrate the ability of the multigrid method to handle
ill-conditioned systems, test calculations were performed on a 32-atom
GaN cell in the zinc-blende structure with the Ga d-electrons in
valence, and on a highly elongated 96-atom diamond cell of dimensions
(6.72, 6.72, 80.64) bohr. For the GaN cell a grid spacing of 0.175
bohr was used, which corresponds to an energy cutoff of 160 Ry in a
plane-wave calculation.  Only the $\Gamma$ point was included in
k-space sampling.  Starting with random initial wavefunctions, 240
self-consistent steps were required to converge the total energy.
Recently, several calculations have been performed on GaN that
explicitly included the d-electrons in valence; the multigrid results
are in good agreement with these calculations (see Table \ref{GaN}).
The calculations for the elongated diamond cell used a grid resolution
of 0.336 bohr, corresponding to an energy cutoff of 35 Ry in a plane
wave calculation.  For convenience, an approximate solution was
generated on a grid with twice the final spacing, on which each
iteration is eight times faster than on the final grid.  After
transferring this approximate solution to the final grid, only 50
self-consistent iterations were sufficient to converge the total
energy to a relative tolerance of $10^{-8}$.

These two ill-conditioned systems represent worst-case scenarios, and
the performance of the multigrid method is considerably better for
typical systems.  Table \ref{ConvergenceResults} illustrates the
convergence properties as a function of grid resolution for an 8-atom
diamond cell.  The observed convergence rates are largely independent
of the energy cutoff.  The number of self-consistent steps required to
converge the density is also nearly independent of the system size.
At an equivalent cutoff of 35 Ry, the multigrid method required 17
steps to converge the total energy of the 8-atom diamond cell to a
tolerance of $10^{-8}$.  When the same calculation was performed on
the 64-atom cell only 20 SCF steps were needed.  This favorable
scaling with respect to both energy cutoff and system size, and the
inherent stability of multigrid methods offer the possibility of
performing electronic-structure calculations on very large systems.

In summary, we have developed a methodology for performing large-scale
{\em ab initio\/} electronic-structure calculations entirely in real
space.  The method uses highly-efficient multigrid techniques to
accelerate convergence rates, which is particularly important for
ill-conditioned systems requiring high energy cutoffs or with large
length scales.  In addition, the multigrid method is readily adaptable
to parallel computer architectures, and its block structure makes it
very suitable for $O(N)$ approaches.
\begin{table}
\caption{{\em Mehrstellen} discretization weights in 3D (h
is the grid spacing).}
\begin{tabular}{crr}
Grid point        & $6h^2$*A & 6*B \\
\tableline
central           & -24      & 6 \\
nearest neighbors &  2       & 1 \\
second neighbors  &  1       & 0 \\
\end{tabular}
\label{Mehrstellen}
\end{table}
\begin{table}
\caption{Comparison of perfect crystal and vacancy in diamond results
obtained using 64-atom supercells [eV].}
\begin{tabular}{lddd}
 &
\multicolumn{1}{c}{Car-Parrinello\tablenote{35 Ry cutoff and $\Gamma$
point sampling.}} &
\multicolumn{1}{c}{Multigrid} &
\multicolumn{1}{c}{Experiment\tablenote{Kittel. \cite{Kittel}}} \\
\tableline
Perfect Crystal \\
\hskip 0.7em Band gap        & 4.53 & 4.53 & 5.50 \\
\hskip 0.7em Cohesive energy & 8.49 & 8.54 & 7.37 \\
Vacancy \\
\hskip 0.7em Formation energy& 6.98 & 7.07 \\
\end{tabular}
\label{Diamond}
\end{table}
\begin{table}
\caption{Comparison of GaN results obtained with Ga 3d electrons in
valence.  See text [eV].}
\begin{tabular}{ldd}
          & Band Gap & Cohesive Energy \\
\tableline
Fiorentini {\em et al.} \tablenote{Full potential LMTO. \cite{FiorentiniEtAl}}
          & 2.00     & 10.89 \\
Wright {\em et al.} \tablenote{Plane wave calculation with 240 Ry
		      cutoff. \cite{WrightNelson}}
          & 1.89 \\
Multigrid \tablenote{32-atom supercell, 160 Ry cutoff and $\Gamma$ point
          sampling.}
          &  1.79    & 10.68 \\
\end{tabular}
\label{GaN}
\end{table}
\begin{table}
\caption{Multigrid convergence tests in an 8-atom supercell:
grid spacing [bohr], equivalent plane-wave cutoff [Ry], total energy
[a.u.], and the number of steps to converge the density.}
\begin{tabular}{cddd}
Grid Spacing & Cutoff & Total Energy & SCF steps \\
\tableline
0.421        & 25.    & -44.87968    & 22 \\
0.336        & 35.    & -45.03331    & 17 \\
0.280        & 60.    & -45.06240    & 21 \\
0.210        & 110.   & -45.06605    & 26 \\
\end{tabular}
\label{ConvergenceResults}
\end{table}

\begin{references}
\bibitem{Cutoff}
   We assume the electronic charge density is expanded with energy cutoff
   twice as large as that of the wavefunctions.
\end{references}
\end{document}